\documentclass[prl,twocolumn, showpacs,amsmath,amssymb]{revtex4}

\usepackage{graphicx}
\usepackage{amssymb}
\usepackage{amsmath}
\usepackage{enumerate}

\begin{document}

\title{Stress clamp experiments on multicellular tumor spheroids}

\author{ Fabien Montel$^{1}$}\author{Morgan Delarue$^{1}$}\author{Jens Elgeti$^1$}\author{Laurent Malaquin$^1$}\author{Markus Basan$^1$}\author{Thomas Risler$^1$}\author {Bernard Cabane$^3$}\author{Danijela Vignjevic$^2$}\author{Jacques Prost$^{1,3}$}\author{Giovanni Cappello$^{1}$}\email{giovanni.cappello@curie.fr}\author{Jean-Fran\c{c}ois
Joanny$^{1}$}
\affiliation{$^1$UMR 168, Institut Curie, Centre de Recherche, 26 rue d'Ulm 75005 Paris France.}
\affiliation{$^2$UMR 144, Institut Curie, Centre de Recherche, 26 rue d'Ulm 75005 Paris France.}
\affiliation{$^3$ESPCI, 10 rue Vauquelin, 75005 Paris, France. }

\date{\today}

\begin{abstract} The precise role of the microenvironment on tumor growth is poorly understood. Whereas the tumor is in constant competition
with the surrounding tissue, little is known about the mechanics of this interaction. Using a novel experimental procedure, we study
quantitatively the effect of an applied mechanical stress on the long-term growth of a spheroid cell aggregate. We observe that a stress of
$10$kPa is sufficient to drastically reduce growth by inhibition of cell proliferation mainly in the core of the spheroid. We compare the
results to a simple numerical model developed to describe the role of mechanics in cancer progression. \end{abstract}

\maketitle

Cancer progression occurs in several stages. In the case of carcinomas, which are cancers of epithelial cells, the primary tumor grows
locally, until some cells invade the neighboring tissue called the stroma which is essentially made of extracellular matrix, fibroblast cells,
immune cells and capillary vessels. Three key elements control proliferation of the primary tumor: the accumulation of gene mutations and the
tumor biochemical and mechanical micro-environments. It is difficult to isolate in vivo one of these factors to measure accurately its
importance. Several recent works suggest that mechanical stress plays a role in tumor progression. A mechanical stress applied to genetically
predisposed tissues or tumor spheroids grown in vitro induces signaling pathways that are characteristic of cancer invasion
\cite{Whitehead2008, Demou2010}. It has also been shown that an increase of mechanical stress leads to a reduction in cancer cell
proliferation in vitro, and drives apoptosis through the mitochondrial pathway \cite{Cheng2009,Roose2003,Helmlinger1997}. In spite of these
experimental evidences, the precise role of the micro-environment, and its interaction with the tumor, are poorly understood. Our group has
developed a theoretical framework \cite{Basan2009,Ranft2010, Basan2011} to describe the influence of the balance between cell division and
apoptosis on tumor growth under stress. The theory is based on the existence of a homeostatic state of a tissue. This is the steady state of
the tissues where cell division balances cell death. The homeostatic stress is a function of the biochemical state of the tissue and depends
on the local concentrations of nutrients, oxygen and growth factors as well as on the environment of the tissue. Signaling induced by the
stroma can for example modify the homeostatic state. In the simple case where the biochemical state of the tissue can be maintained constant,
the homeostatic stress is the stress that the tissue can exert at steady state on the walls of a confining chamber. It is a measure of
mechanical forces that cells can sustain in this state. Indeed to grow against the surrounding tissue, cells have to exert mechanical stress
on the neighboring cells.

In this paper we test experimentally the relevance of the homeostatic stress concept. We measure the effect of a known external stress on the
growth of a cellular aggregate mimicking a tumor over timescales longer than the typical time scales of cell division or apoptosis. We use a
new experimental strategy to exert a well defined mechanical stress on multicellular tumor spheroids for a period of time exceeding 20 days.
 
We prepare colon carcinoma cell spheroids derived from mouse CT26 cell lines (ATCC CRL-2638) using a classical agarose cushion protocol
\cite{Hirschhaeuser2010}. The wells of a 48 wells plate are covered with agarose gel (Ultrapure agarose, Invitrogen Co, Carlsbad, CA) and cell
suspensions are seeded on the gels at concentration of $20 000$ cells per well. Cells self-assemble into spheroids in less than 24h. Cells are
cultured under 95\% air/ 5\% CO$_2$ atmosphere in DMEM enriched with 10\% calf serum (culture medium). Using confocal microscopy, we check
that the shape of the spheroid is indeed close to a sphere. A constant stress is applied on the tissue over long time scales by imposing the
osmotic pressure of a solution of the bio-compatible polymer Dextran ($M_w=100$kDa, Sigma-Aldrich Co, St Louis, Mo). This polymer is known to
be neutral and is not metabolized by mammalians cells. We also confirmed that it is neither a growth or a death factor by plating cells for 3
days with Dextran and measuring cell concentration and viability.

\begin{figure} \centering \includegraphics[height=9cm]{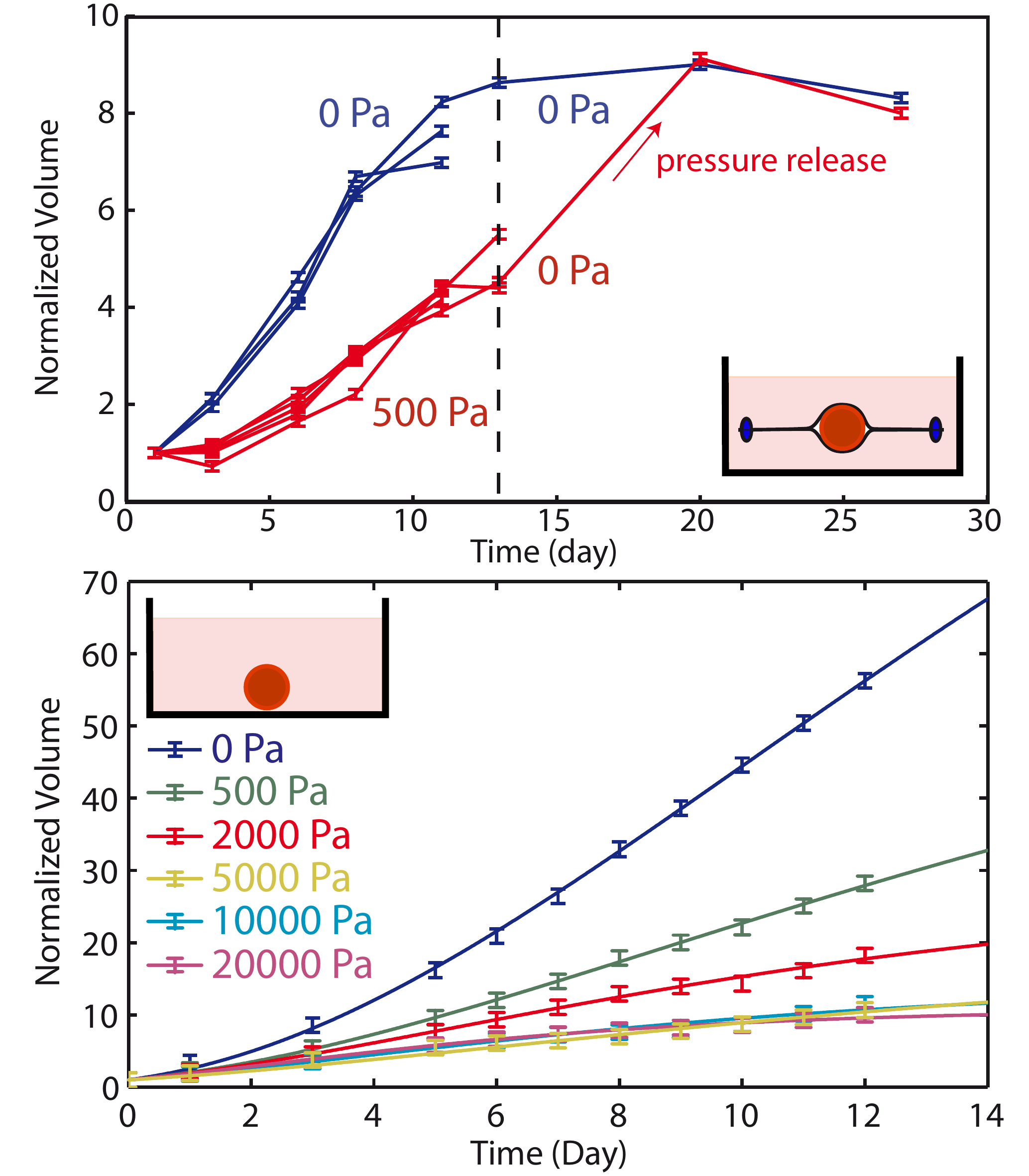} \caption{Growth curves of individual spheroids under stress. (Top)
Normalized volume of individual spheroids as a function of time for the indirect experiments. The initial diameter is $D_o = 350 \mu m$. At $t
= 12$ days, stress is released. (Bottom) Normalized volume of individual spheroids as a function of time for the direct experiments. The
initial diameter is $D_o = 200 \mu m$. The inserts show the principles of the two experiments. The points are representative experimental data
for individual spheroids taken out of a larger set of experiments. For each condition, $N\geq3$ experiments have been recorded. The lines are
the results of fits with the two rate model. Error bars are the image analysis errors.} \label{Growth curves} \end{figure}

We first perform {\em indirect stress measurements}. A growing spheroid is positioned inside a closed dialysis bag (diameter 10 mm,
Sigma-Aldrich) which is then placed in an external medium with added Dextran. The dialysis membrane was chosen so that its molecular weight
cut-off (10 kDa) impedes the diffusion of Dextran. The osmotic stress induces a force on the dialysis membrane, which is transmitted in a
quasi-static equilibrium to the spheroid and calibrated as in \cite{Belloni1994, Bouchoux2009}. The stress exerted on the cellular system can
be seen as a network stress that tends to reduce the volume occupied by the spheroid. It acts directly on the cells, and not on the
interstitial fluid. The volume $V(t)/V_0$, normalized by the initial volume of the spheroid $V_0$, is measured at successive times from a top
view using differential interference contrast microscopy (Axiovert 100, Zeiss). In the absence of any applied stress, the spheroid reaches a
steady state with a typical diameter 900 $\mu$m. When Dextran is added to the medium, a decrease of the growth rate $\frac{dV}{dt}$ and of the
steady state volume are observed (Fig.\ref{Growth curves}-Top). Interestingly, after a stress release, the growth of the spheroid resumes
until it reaches the same steady state volume as in the absence of external pressure. This indicates that the effect of stress is fully
reversible. Altogether these results show that an external applied stress modulates the growth of tumor spheroids.

We have also performed {\em direct experiments} where the osmotic stress is applied onto the spheroid in the absence of the dialysis membrane.
In order to verify that Dextran cannot diffuse inside the spheroid, we have placed it in a medium supplemented with fluorescent FITC-Dextran
at an osmotic stress $\Pi=1000$Pa. After 4 days of incubation, the first 70 $\mu$m of the spheroid were imaged using spinning disc microscopy.
We measured that the amount of Dextran able to penetrate into the spheroid is negligible compared to the Dextran concentration in the medium.
The osmotic stress is thus applied on the first layer of cells that plays the role of the dialysis membrane in the {\em direct} experiment and
transmits the stress to the rest of the spheroid. The volume of the spheroid has also been measured as a function of time (Fig.\ref{Growth
curves}-Bottom). We observe a dependence of the growth rate and the steady state size on stress very similar to that observed in the {\em
indirect} experiment, validating our approach. Interestingly, for a stress larger than $10 kPa$ the effect of stress saturates and the growth
curves are indistinguishable from each other.

\begin{figure} \centering \includegraphics[height=5.8cm]{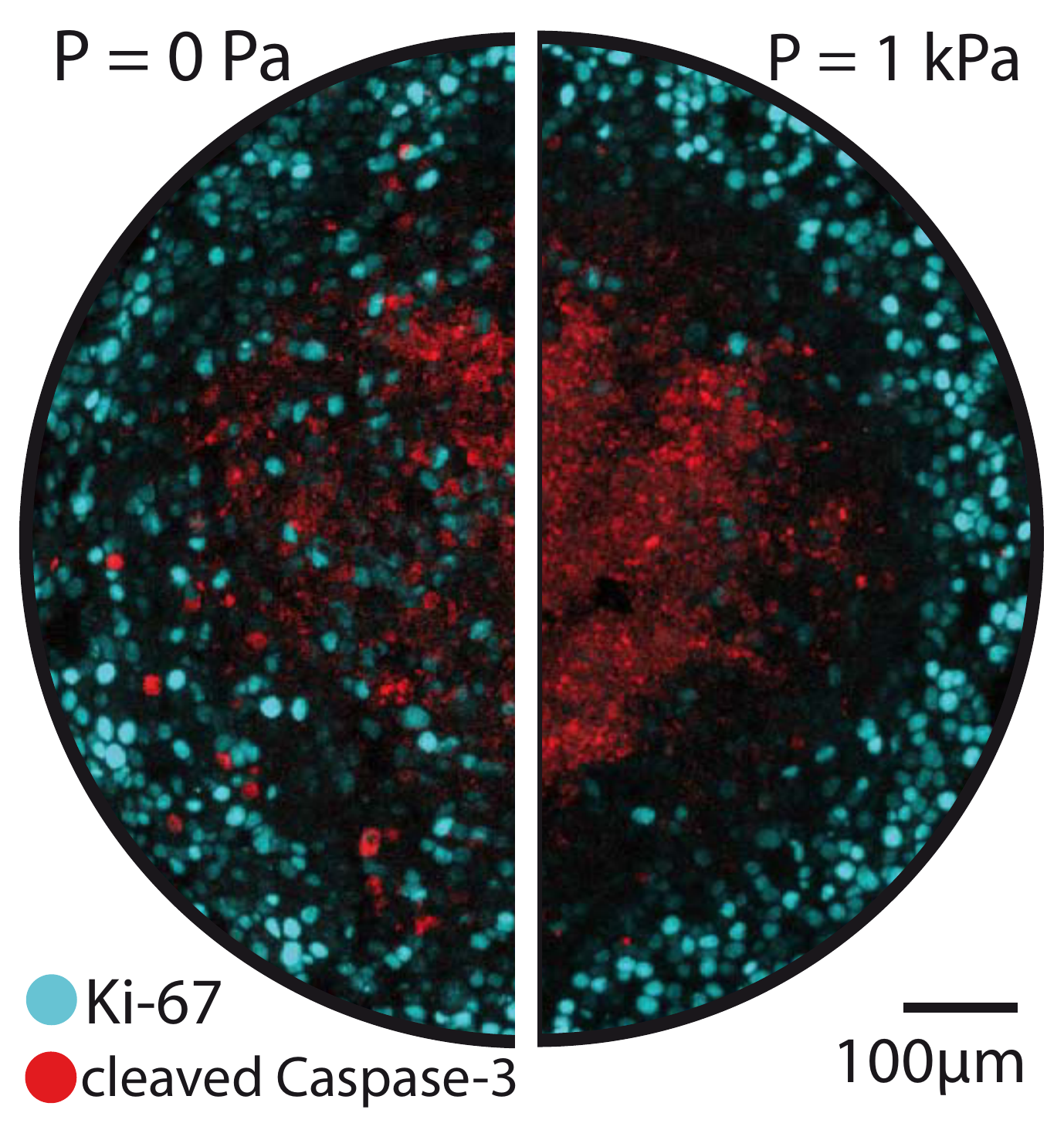} \caption{Effect of stress on the distribution of proliferation and
apoptosis. Cryosections and immunofluorescence of the spheroids are used to label the cell divisions (antibody against Ki67 in cyan) and
apoptosis (antibody against Cleaved-capase3 in red).(Left) Half section of a spheroid grown in normal medium for 4 days (Right) Half section
of a spheroid grown with a stress of $1$ kPa for 4 days} \label{Cryo} \end{figure}

The direct experiment is based on the application of a mechanical stress on the surface of the spheroid through an osmotic shock. Osmotic
stress is known to have direct effects on cell growth and apoptosis in particular through the mitogen activated protein kinase (MAPK) pathway
\cite{Cowan2003, Racz2007, Nielsen2008, Xie2007}. However, in all these studies, the effect of an osmotic shock is only measured for an
osmotic stress two orders of magnitude larger than the one applied in our experiments ($1 MPa$ compared to $10 kPa$). Moreover we do not
observe any apoptosis at the surface of the spheroid where the osmotic stress is exerted (Fig. \ref{Cryo}). In addition, it can be checked by
balancing chemical potentials that the presence of Dextran outside the spheroid creates a negligible concentration gradient of all other
solubles molecules. The concentration difference of a small soluble solute exchanged between the interior and the exterior of the spheroid can
be estimated as $\Delta c_s / c_s = v\Pi/kT (1-\Theta_s)<10^{-3}$ where $c_s$ and $\Theta_s$ are the concentration and the volume fraction of
the solute, $v$ the molecular volume of the solvent and $\Pi$ the applied osmotic stress. In other words, the chemical potential of water in
the cell is dominated by the small ions and it is only slightly modified by the presence of Dextran.

Finally we investigate experimentally the spatial dependence of cell division and apoptosis using cryo-sections and immuno-fluorescence.
Spheroids of comparable diameters are embedded in a freezing medium, placed at $-80^{o}$C and cut in slices of 5 $\mu$m thickness at the level
of their equatorial plane. Using a classical immuno-staining protocol, we then label fluorescently dividing cells (in cyan with an anti-Ki67
antibody) and apoptotic cells (in red with an anti -cleaved caspase3 antibody) (See Fig. \ref{Cryo}). We observe that in the absence of
external stress, cell division is distributed over all the spheroid with an increase at the periphery, whereas for an external stress of
$1$kPa, it is greatly reduced in the center of the sections. As in previous studies \cite{Mueller-Klieser1997, Hirschhaeuser2010}, we observe
an accumulation of apoptotic cells in the center of the spheroid but with no measurable effect of stress on this localization.

In order to better understand this stress dependence of cell division and to interpret the generic trends of the experimental findings, we
performed numerical simulations similar to those of Ref. \cite{Basan2011, Basan2009}. We adapt these simulations to the geometry and setup of
the experiments. In brief, in the simulations, a cell is represented by a pair of particles which repel each other and thus move apart. When a
critical distance is reached, the cell divides. After division, each original particle constitutes, together with a newly inserted particle in
its surrounding, a daughter cell. Particles belonging to different cells interact with all particles with a short range interaction: a
constant attractive force describes cell-cell adhesion, while a repulsive short range potential ensures volume exclusion. The viscous drag
between cells is taken into account by a ``Dissipative Particle Dynamics`` -type thermostat. Finally a constant apoptosis rate provides cell
removal. In order to mimic the experiments, tissue spheroids are grown in a container together with a ''passive liquid'' under stress. This
liquid interacts with the cell particles in a similar way as cell particles with each others and it transmits the stress on the spheroid.

\begin{figure} \centering \includegraphics[width=8.5cm]{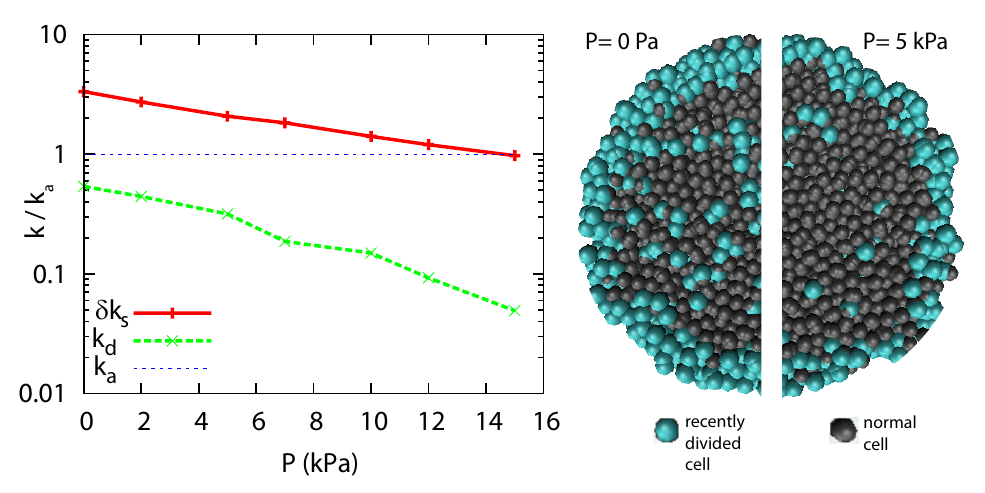} \caption{Dissipative Particle Dynamics Simulations. (Left) Bulk
division rate $k_d$, surface rate increment $\delta k_s$ and apoptosis rates $k_a$ as function of the applied pressure $P$. (Right) Virtual
cryosections of the simulated spheroids for an external pressure P=0 Pa or P=5kPa. The pressure units have been calibrated from the
experiments using the saturation pressure.} \label{Simulations} \end{figure}

As in the experiments, we observe a steady state that depends on the applied stress. In the numerical simulations each division event can be
traced and the spatial distribution of divisions can be measured to build virtual cryosections of the simulated spheroids. We observe a strong
dependence of the division rate on the distance to the surface of the spheroids and a clear decrease of cell division anywhere in the section
but stronger in the core (see Fig.\ref{Simulations}).

Based on the growth curves and cryosection observations, we present a simple two rate description of the spheroid growth in the absence or the
presence of external stress: the core of the spheroid is mostly undergoing apoptosis whereas its periphery is proliferating. In this
situation, the net growth rate is proportional to the area $(\propto r^2)$ while the net death rate is proportional to the volume $(\propto
r^3)$. This surface growth effect leads to a stable steady state size. The surface localization of the proliferation can be obtained using
purely mechanical considerations. A cell must deform its environment to grow. The deformation is facilitated if the cell is closer to the
surface, and this implies that proliferation is favored at the surface. The increased number of cell divisions at the surface drives a flow
from the surface of the spheroid toward its center. The flow is a possible explanation for the accumulation of the long lasting apoptotic
markers in the center of the spheroid. A mechanical control of cell cycle entrance can also explain the growth of tumor spheroids in free
suspension \cite{Drasdo2005a}. In this case nutrient depletion causes the formation of a necrotic core which generates death at the inner
surface of the viable rim $(\propto r^2)$ i.e. not proportional to the volume thereby not generating a steady state \cite{Drasdo2005a,
Schaller2005}. Using a fluorescently labelled growth factor (Alexa 555- EGF) we have verified that the transport of these molecules is not
affected by stress. This result supports a direct mechanical effect on the division rate.

Our two rate model can be seen as a simplified version of the two rate model of Radszuweit et al. \cite{Radszuweit2009}. The net bulk growth
rate is $ k=k_d-k_a$, where $k_d$ and $k_a$ are the division and apoptosis rates respectively. It is a function of stress. At the surface, the
net growth rate $k_d-k_a+\delta k_s$ is larger and $\delta k_s$ has a different stress dependence. Taking into account surface and bulk
growth, the growth equation reads : \begin{equation} \label{eq:growth} \partial_t N=(k_d -k_a)N+\delta k_s N_s \end{equation} Assuming a
constant cell density and a constant thickness $\lambda$ of the region where the division rate increment is equal to $\delta k_s$, one can
express for $R>\lambda$ the rate of volume increase as : \begin{equation} \label{eq:growth2} \partial_t V=(k_d -k_a)V+ (36\pi)^{1/3}\delta
k_s\lambda V^{2/3} \end{equation}

\begin{figure} \centering \includegraphics[height=4cm]{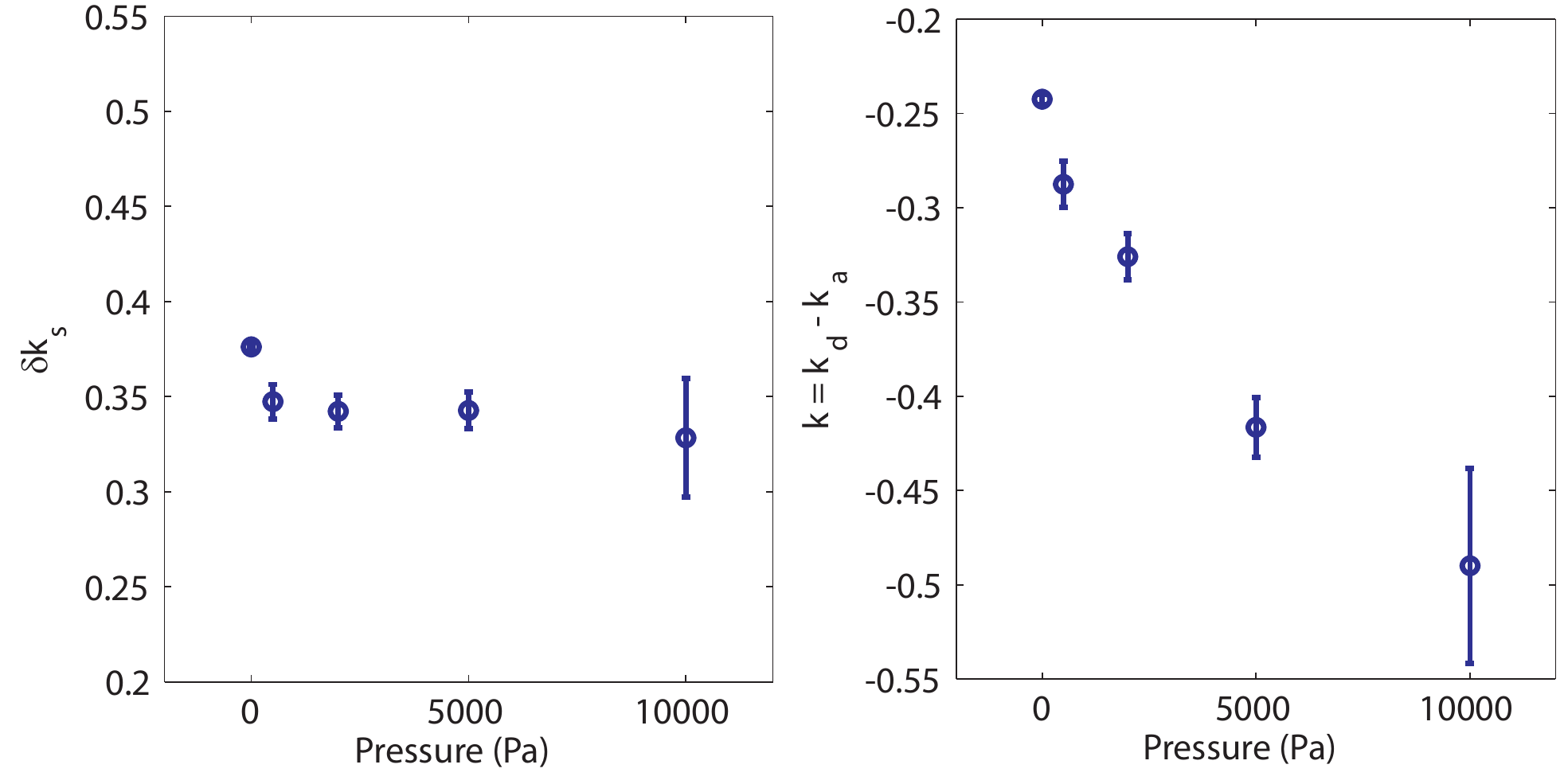} \caption{Evolution of growth rates with stress. (Left) Surface
division rate increment $\delta k_s$ as a function of stress. (Right) Bulk growth rate $k$ as a function of stress. For each condition,
$N\geq3$ experiments have been recorded. The errors bar are obtained using a jackknifing method and represent the efficiency of the fitting
algorithm} \label{Surface model} \end{figure}

For small spheroids ($R<<\lambda$) the growth rate is positive and constant ($k_d+\delta k_d-k_a$), this leads to the previously described
exponential growth \cite{Drasdo2005a, Schaller2005}. In our case the growth curves can readily be fitted by Eq. \ref{eq:growth2} in the range
of large spheroids \cite{bertal1957}. The variation with pressure of the parameters $k = k_d - k_a$ and $\delta k_s$ is given in Fig.
\ref{Surface model}. The surface growth rate $\delta k_s$ is less affected by stress than the bulk growth rate $k$. A similar fit can be
performed on the simulations. The bulk and surface growth rates $k$ and $\delta k_s$ are represented on Fig. \ref{Simulations}. Although both
the surface and bulk growth rates depend exponentially on pressure, the decay constant of the surface rate is much smaller than that of the
bulk rate. While the bulk rate decreases by more than one order of magnitude, the surface rate decreases by a factor 3. This supports the
hypotheses of the simulations. In summary, cell division in the core of the spheroid is strongly affected by stress whereas cell division rate
increment on the surface of the spheroid depends more weakly on stress.

In conclusion, we have shown by a direct measurement of the tissue response to an external stress that the application of an external stress
drastically limits the growth of tumoral spheroids. Previous approaches \cite{Helmlinger1997, Fritsch2010a} had used the elastic deformation
of poro-elastic gels to measure the maximum stress that can be developed by spheroids. The measured stress in these experiments is in the same
range as in our measurements. In a recent study a localized increase of mitochondrial apoptosis and a reduction of proliferation in presence
of stress were also reported \cite{Cheng2009}. This difference with our results may be due to the fact that in our case we are not controlling
the rigidity of the surrounding substrate but the applied pressure. This may leads to a different response of the spheroid. Our results favors
the idea that direct mechanical effects can have strong implications in cancer proliferation. This raises the question of the players in the
crosstalk between stress and cellular response and in particular of the nature of the stress sensor.

We would like to thank F. Brochard, F. Graner, P. Nassoy, K. Alessandri and J. Kaes for useful discussions. F.M. and G.C. would like to thank
Axa Research Fund and CNRS for funding. The group belongs to the CNRS consortium CellTiss.


\end{document}